# AI Debris: Residual Risk and the Afterlife of Failed AI Systems

**Victor Frimpong**[1]

**Abstract:** AI governance frameworks primarily focus on risks during the development and deployment phases, implicitly treating system withdrawal as merely a technical shutdown. This paper argues that decommissioned AI systems generate residual risk—"AI debris"—that persists after models are removed and continues to shape institutional behaviour, accountability, and trust. AI debris is defined as the post-withdrawal socio-technical residue of AI systems, including workflow dependency, data contamination, capability displacement (deskilling), legitimacy erosion, and accountability breakdown, and is distinguished from adjacent concepts such as technical debt and safety-critical post-incident review on the grounds of temporal locus and institutional scope. The paper develops a typology of six debris domains and explains the mechanisms by which debris persists in institutional memory, path dependency, blame avoidance, and feedback effects within organisational data. To operationalise the concept, it proposes an evaluator-ready AI Debris Decommissioning Protocol (AIDP): a stepwise checklist specifying auditable evidence for freezing decision footprints, incident review, remediation, contestability, and post-withdrawal accountability assignment, alongside a discussion of the incentive conditions—external enforcement through procurement or regulation, and protected disclosure—under which such a protocol is plausibly adopted. Two illustrative cases, Amazon's discontinued hiring tool and the Dutch childcare benefits scandal, demonstrate how algorithmic systems can institutionalise decision categories, screening heuristics, and unresolved harm that persist well after formal withdrawal. The paper contributes a practical governance instrument for regulators, auditors, and organisations seeking to prevent "paper compliance," strengthen AI lifecycle governance, and improve institutional resilience in high-stakes decision environments.

**Keywords**: AI governance, AI lifecycle, residual risk, decommissioning, accountability, socio-technical systems, institutional resilience

**JEL:** O33, D83, G38

## 1. Introduction: Why AI remains relevant even after decommissioning

Research on AI governance has grown rapidly, driven by the widespread adoption of algorithmic decision-making systems across public administration, banking, healthcare, education, and labour markets. This body of work has mainly focused on mitigating risk during active deployment through fairness evaluations, transparency, explainability, risk classification, and accountability mechanisms (Barocas et al., 2019; Kroll, 2017; Raji et al., 2020).

[1] SBS Swiss Business School, Flughafenstrasse 3, 8302 Kloten-Zürich, Switzerland, Email: v.frimpong@research.sbs.edu

An important institutional reality remains underexplored: many AI systems do not endure. They may be paused, replaced, quietly discarded, or withdrawn due to controversies, performance issues, political backlash, legal challenges, or reputational damage. For companies, AI-generated residuals pose operational risks and governance concerns that affect reputational equity, regulatory vulnerability, and future AI adoption.

In practice, organisations often consider AI withdrawal a final step. Removing these systems seldom restores the institutional conditions that existed before implementation (Sculley et al., 2015). Decision-making processes may remain altered, datasets may remain tainted, trust may continue to erode, and accountability may remain unresolved. Conventional frames of "AI risk," which generally focus on immediate, quantifiable outcomes, do not adequately address these lingering effects (Kondor et al., 2024). What is lacking is a framework for understanding the aftermath of AI systems—what persists when they are no longer functional.

The paper argues that AI governance should extend beyond deployment to include a regulated phase. It introduces the concept of AI debris: the residual socio-technical and institutional risks and effects that persist beyond the system's active use. The proposition is straightforward but relevant to governance: while AI systems are discontinued, their institutional traces linger. The paper provides (i) a definition of AI debris, (ii) a classification of six types of debris, and (iii) a governance recommendation: AI governance should broaden its scope beyond ex ante compliance and in-deployment monitoring to include decommissioning protocols, remediation obligations, and post-incident institutional repair.

## 2. AI governance is deployment-centric

A significant portion of modern AI governance has been influenced by the necessity to oversee AI systems once they are operational (Laux, 2023). This includes algorithmic audits (Raji et al., 2020), accountability frameworks for automated decision-making systems (Kroll, 2017), and a growing array of fairness evaluation techniques and normative discussions (Barocas et al., 2019; Binns, 2018). These strategies are crucial, yet inherently centred on deployment: they presuppose that the AI system is operational and can be assessed, modified, monitored, and enhanced.

An increasing number of actual AI systems operate under conditions less stable than governance frameworks might suggest. Systems can be discontinued for various reasons: political issues, vendor withdrawal, regulatory actions, public backlash, notable incidents, or internal cost-effectiveness reassessments. However, discontinuation does not remove governance responsibilities. Instead, it marks the beginning of a new governance stage: the post-deployment phase.

This gap is significant because the repercussions of AI failure are often deferred, dispersed, and mediated through institutions. A system might be taken out of service, but the organisation still suffer a loss of legitimacy, internal disarray, and degraded capabilities. The governance challenge extends beyond assessing model fairness; it also encompasses the consequences for the institution after a model is deployed and then withdrawn. As with systemic risk, harms after deployment can be distributed, delayed, and shaped by organisational structures, making it difficult to link them to a specific technical failure. Many governance frameworks view AI as a lifecycle-managed system, but they primarily focus risk controls on development and deployment (Raji et al., 2020). This leaves system withdrawal and post-decommission residual risk less defined (NIST, 2023).

Two areas of research provide a robust basis for understanding this. First, sociotechnical views emphasise that technology is never solely technical; it is situated within institutions, norms, and structures of authority (Latour, 2005). Second, legitimacy studies show that organisational legitimacy is not easily regained through remedial measures; it is socially constructed and often path-dependent (Suchman, 1995). These fields of literature suggest that the aftermath of AI systems is not incidental—it is predictably structured.

Current AI governance frameworks generally focus on managing risks during the deployment phase and treat shutdown as merely a technical process. This perspective fails to account for a post-deployment governance stage in which institutional risk persists even after the systems are taken offline. The concept of AI debris addresses this gap by identifying the enduring impacts of AI systems and formalising decommissioning as a governance responsibility rather than merely an administrative outcome.

AI debris differs from technical debt and its extension to machine learning (Sculley et al., 2015), which involves development shortcuts that create long-term maintenance costs. Technical debt is a future-oriented, in-service issue, resolved through refactoring while systems are active (Kruchten et al., 2012). AI debris, however, is a post-service issue—what remains in institutions after the system is gone, when refactoring is impossible. While engineers manage technical debt on live systems, AI debris requires organisations to address residual impacts without the system.

The second literature area is decommissioning and post-incident review in safety-critical fields like aviation and medical devices, where structured processes (e.g., root-cause analysis, after-action reviews, incident reporting) oversee system retirement and failure investigation. AI debris shares the governance premise of decommissioning but differs in scope and mechanism. Safety protocols focus on small, physically bounded systems with clear failure signatures and centralised oversight (e.g., FAA, FDA) (Abulibdeh et al., 2025). AI systems, by contrast, span decision workflows, data pipelines, and human judgment; they are not physically bounded or easily instrumented, and sectoral regulators do not track their retirement. Safety frameworks mainly address causal failure mechanisms, while AI debris is largely institutional and interpretive, tied to categories, habits, trust, and data that persist beyond a single failure event (Arslan, (2026). AI debris does not replace existing literature but highlights a gap, addressing residual risk that is neither a system maintenance issue (technical debt) nor a bounded, causally traceable failure (post-incident review), but an ongoing institutional condition after the system and investigation conclude.

## 3. Definition

This paper defines AI debris as follows:

> *AI debris refers to the enduring institutional, social, and technical remnants generated by AI systems that persist after their active use and continue to influence decision-making, legitimacy, and organisational capability.*

AI debris can be measured, categorised, and managed. It can be assessed, traced to specific institutions, and handled through post-decommissioning governance. AI systems do not just cease to exist when they're shut down; they often leave behind lasting impacts. Figure 1 illustrates this process, showing the stages from deployment to withdrawal, the accumulation of residual effects, and the lasting institutional consequences.

**Figure 1.** The AI Debris Pathway: From Deployment to Institutional Afterlife

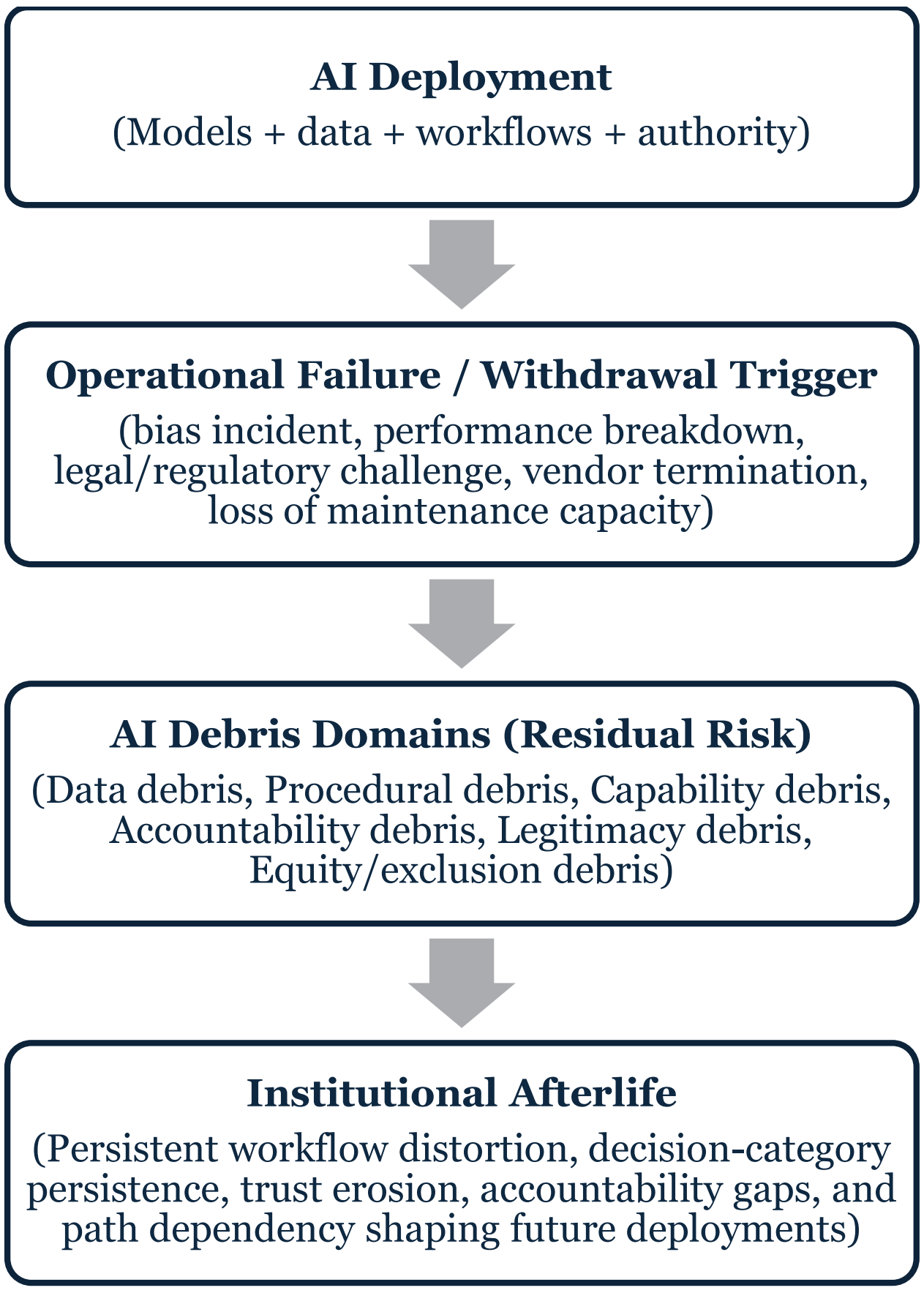

The figure illustrates AI withdrawal as a governance stage rather than merely a technical endpoint. Even after decommissioning a system, lingering risk can continue to affect various areas, influencing institutional behaviour, legitimacy, and future decision-making frameworks. AI debris becomes most apparent when systems fail or are removed under scrutiny, but it can also build up in successful implementations due to procedural reliance, data issues, and shifts in capabilities (Eubanks, 2018; O'Neil, 2016). Unlike traditional IT systems, AI systems create decision categories, risk scores, and established authority structures that are often perceived as objective (Kroll, 2017; Ananny & Crawford, 2018). AI debris differs from "AI risk" in three key respects.

1. AI debris is temporal and delayed: AI-related risk often remains hidden. For instance, an AI hiring tool may be discontinued due to bias, but the organisation's hiring process can still be influenced by AI-generated screening criteria. Thus, the issue lies not in the model itself, but in the ongoing institutional structures that persist after its removal.
2. AI debris is often invisible: AI debris often appears as leftover processes, data errors, and diminished capabilities. Standard evaluation methods often overlook these issues because they focus on measurable outcomes such as error rates, disparities, or performance standards (Barocas et al., 2019).
3. AI debris is governance-relevant: AI debris is a critical governance issue, impacting responsibility allocation, institutional learning, and legitimacy repair (Kroll, 2017; Suchman, 1995). Organisations can meet governance requirements yet still create debris that weakens institutional resilience.

*3.1 Illustrative case 1: Amazon's discontinued hiring tool*

Between 2014 and 2017, Amazon developed an experimental machine-learning recruiting tool that scored job applicants on a scale from one to five stars, based on patterns extracted from a decade of previous hiring decisions. Due to the fact that the majority of those historical hires had been men, particularly in technical roles, the model learned to penalise résumés containing the term "women's" and to downgrade graduates of two women's colleges, thereby effectively instructing itself that male candidates were preferable (Dastin, 2018). Amazon discontinued the project after determining it could not reliably correct the model's gender bias; reports indicate recruiters consulted the tool's recommendations without relying solely on them, and that a reduced version of the system remained in limited use into 2018 (Dastin, 2018). The system was never subjected to an independent audit, and no organisational post-mortem has been publicly disclosed. Consequently, publicly available information pertains to the trigger for withdrawal rather than its institutional aftermath: no independently verified account shows how Amazon's subsequent hiring workflows, résumé-screening heuristics, or recruiter judgment were influenced by three years of exposure to the tool's outputs. Any assertion regarding the tool's persistent influence on Amazon's human resources practices is, therefore, an inference based on mechanisms rather than documented evidence—a limitation that this paper considers a data point in its own right, given that a fully deployment-focused reporting environment that provides detailed information on bias incidents but virtually nothing on organisational consequences exemplifies the scarcity of evidence once a system is decommissioned.

This case is employed descriptively to support the plausibility of what this paper terms procedural, capability, and equity debris—not as definitive evidence that such outcomes actually occurred at Amazon.

### 3.2 *Illustrative case 2: the Dutch childcare benefits scandal*

From 2013 to 2019, the Dutch Tax and Customs Administration (Belastingdienst) used a self-learning risk classification algorithm to flag childcare benefit applications that could be fraudulent. This system incorporated dual nationality as a risk factor and disproportionately targeted families with a migration background (Amnesty International, 2021). In contrast to the Amazon case, the subsequent consequences of the system's discontinuation are documented not through journalistic inference but via an official state investigation. A parliamentary inquiry committee concluded in its report, 'Ongekend Onrecht' ("Unprecedented Injustice'), that approximately 26,000 families had been erroneously accused of fraud, often compelled to reimburse sums amounting to tens of thousands of euros, and that the conduct of the agency violated fundamental principles of the rule of law (Tweede Kamer, 2020). The termination of the system did not end the episode; rather, it initiated a multi-year institutional aftermath that continues to affect several domains.

Equity debris is apparent in the downstream consequences of the scandal: affected families were driven into debt and bankruptcy, and in more than 1,600 instances, children were removed from their homes by youth-protection services as a secondary result of the financial and administrative repercussions (Amnesty International, 2021). Legitimacy debris is observable at the institutional level: the report's publication prompted the incumbent cabinet's resignation in January 2021, showing how algorithmic failure can incur legitimacy costs that persist for years beyond the system's lifespan. Accountability debris is evident in the subsequent compensation process: government reports in 2021 indicated that redress payments had been made to fewer than half of recognised victims, showing that formal attribution of responsibility for a withdrawn system does not, in isolation, resolve operational issues.

While the Amazon case shows what AI debris might look like, the Dutch case provides it with an auditable evidentiary record—such as parliamentary inquiry findings, government redress statistics, and independent human rights investigations—that the AIDP (Table 1) aims to ensure in the future.

AI debris persists even after a system is withdrawn for several reasons. First, institutional memory can sustain AI-generated categories, thresholds, and decision-making processes as informal "best practices" even after the system's removal. Second, data contamination occurs when AI outputs are reintroduced into organisational records, thereby distorting classifications and influencing future decisions. Third, path dependency arises as workflows and performance expectations adapt to the AI system, making it difficult to revert to pre-AI methods. Fourth, blame avoidance can lead individuals to retain AI-driven justifications, obscuring accountability after decommissioning. Finally, repeated reliance on AI can erode human judgment, making institutions less effective once the system is gone. These factors collectively explain why AI's negative impacts can persist long after its technical shutdown (Bainbridge, 1983; Jasanoff, 2004; O'Neil, 2016).

## 4. A Typology of AI Debris

To operationalise the concept, this research note proposes six analytically distinct types of AI debris. These categories can be separated, but they frequently interact in real-world situations.

### *4.1 Legitimacy debris*

Legitimacy debris refers to the lasting damage to an organisation's credibility, trust, and acceptance, even after an AI system is removed. Organisational legitimacy is not binary; it's socially granted and upheld through signals, but undermined by failures (Suchman, 1995). When authoritative AI systems fail, they can create legitimacy shocks. Even if the issues are resolved, or the system is taken down, legitimacy does not automatically recover.

This issue is critical in public-sector and high-stakes environments, where AI is often promoted as a means of modernisation, efficiency, or fairness. If real experiences contradict these claims, institutions face enduring trust deficits. Legitimacy debris symbolises a form of path dependency—damaged trust affects how future reforms are viewed, leading to scepticism (Suchman, 1995).

### *4.2 Accountability debris*

Accountability debris refers to unresolved responsibilities that remain after a system is withdrawn. In AI governance, accountability is typically managed through documentation, audits, and compliance. However, responsibility is often fragmented among vendors, developers, managers, data suppliers, regulators, and decision-makers (Kroll, 2017).

When a system fails and is taken offline, organisations avoid assigning blame to sidestep legal risks or reputational damage. This leads to governance residue, in which the institution cannot assign responsibility, learn from failures, or prevent their recurrence. Thus, accountability debris indicates a breakdown in institutional learning, not a flaw in technical design (Kroll, 2017).

Research on algorithmic auditing supports this view: while audits can uncover failures, without proper accountability mechanisms, they become merely symbolic (Raji et al., 2020). This ties into the broader issue of governance performativity.

*4.3 Capability debris*

Capability debris refers to the loss of human judgment, expertise, and decision-making capacity resulting from AI adoption. Research shows that automation can deskill operators, create dependency, and hinder effective intervention during failures (Bainbridge, 1983).

The "ironies of automation" point out that removing humans from routine tasks can leave them ill-equipped for necessary interventions (Bainbridge, 1983). Studies on automation misuse and overreliance indicate that people can develop habits that diminish their independent judgment (Parasuraman & Riley, 1997). Additionally, humans may place excessive trust in machine outputs, even when they are incorrect (Skitka et al., 1999).

When an AI system is removed, organisations often believe that capabilities will naturally return. However, capability debris suggests that the organisation may struggle with a shortage of skilled staff, the loss of tacit knowledge, and diminished confidence in human decision-making (Rinta-Kahila et al., 2018). The absence of the system does not restore the human capacity it displaced.

*4.4 Data debris*

Data debris refers to the lasting effects of corrupted, biased, or mislabeled datasets and feedback loops that arise during AI deployment. AI systems generate data traces, such as decisions, classifications, risk scores, and behavioural categories, which are often re-entered into organisational databases as if they were accurate.

Research on fairness has shown that biased training data can perpetuate structural inequality (Barocas et al., 2019). However, data debris highlights the issue of persistence: even after an AI system is removed, the distortions it creates persist in the data. For example, a biased automated risk-scoring system produces skewed classifications in client records that continue to affect future decisions.

This relates to the "abstraction traps" critique, in which AI simplifies complex social realities into computational categories that are institutionalised as objective (Selbst et al., 2019). Thus, data debris is a mechanism by which these simplifications can become entrenched.

*4.5 Procedural debris*

Procedural debris refers to the lingering effects of AI-driven workflows, routines, and decision-making processes even after removing an AI system. Implementing AI is rarely just a technical change; it involves redesigning workflows, establishing new decision hierarchies, and altering organisational assumptions about decision-making.

From a sociotechnical perspective, technology shapes how institutions are structured, thereby influencing what is considered evidence, authority, and rationality (Latour, 2005). When an AI system is taken out, the underlying workflow often remains intact. Organisations may continue to operate as if AI outputs still influence decisions, using categories, checklists, escalation pathways, or automated decision logic embedded in their processes.

Procedural debris can be particularly problematic because it's not always apparent. Organisations may believe they have resolved issues by removing the AI, yet they continue to rely on AI-shaped decision-making frameworks. Procedural debris is the ongoing impact of AI-driven workflows, while capability debris is the decline in human judgment, even when workflows return to their original state.

*4.6 Equity debris*

Equity debris refers to the lasting exclusion faced by disadvantaged groups even after removing an AI system. Research has shown that automated systems can worsen structural inequality through classification and decision-making (Benjamin, 2019; Eubanks, 2018; O'Neil, 2016). The negative impacts often persist beyond the system's withdrawal.

Equity debris manifests as ongoing disadvantages such as denied access, reputational harm, bureaucratic delays, and the reinforcement of exclusionary categories. For instance, if an automated system wrongly denies someone a benefit, the repercussions can include lost time, income, housing instability, and diminished trust in institutions, all of which cannot be reversed simply by discontinuing the system (Eubanks, 2018).

This illustrates a significant governance failure: institutions often view withdrawal as a solution, yet affected groups continue to face irreversible risks.

The six debris domains are analytically distinct but frequently co-occur in practice, as the illustrative cases in Section 3 show. To make the typology usable by auditors and evaluators—rather than a purely descriptive vocabulary—each domain requires a discriminating question that can, in principle, be answered from observable evidence. Table 1 sets out this discriminant criterion for each domain, distinguishing it from the domain it is most often confused with in practice.

**Table 1** Discriminant criteria for classifying AI debris observations

| Domain | What persists | Discriminating question | Typical evidence | Distinguish from |
|---|---|---|---|---|
| **Legitimacy debris** | Generalised trust deficit | Do outside stakeholders remain sceptical of the institution in general, independent of whether they were personally affected? | Stakeholder/media sentiment, programme uptake decline, trust indices | *Equity:* legitimacy is diffuse and reputational; equity is concrete and borne by identifiable affected parties. |
| **Accountability debris** | Unowned responsibility for remediation | Can the institution name a specific responsible party for remediation today? | No named responsible officer, unresolved complaint backlog | *Legitimacy:* accountability is internal ownership; legitimacy is external perception. An institution can have one without the other. |
| **Capability debris** | Deskilling, loss of tacit competence | Could staff perform the pre-AI task competently today, without the system? | Training-gap assessments, error rates under manual override | *Procedural:* capability concerns who can act competently; procedural concerns what steps are followed, regardless of who executes them. |
| **Data debris** | Contaminated or mislabelled records | Do current records still contain classifications originally produced by the withdrawn system? | Data lineage audits, contamination assessments | *Procedural:* data debris is about record content; procedural debris is about decision process. |
| **Procedural debris** | Workflow logic assuming the system's outputs | Does the workflow still assume the AI's outputs exist, despite its removal? | SOP comparison pre/post withdrawal, override-practice review | *Data and capability debris* (see above). |
| **Equity debris** | Unremedied harm to identifiable parties | Are specific individuals or groups still bearing unresolved harm from the system's past decisions? | Complaint backlogs, redress case records | *Legitimacy* (see above). |

An observation can implicate more than one domain simultaneously. The table lets an evaluator assign a primary domain per observed instance of debris, with secondary domains noted where applicable, rather than forcing a mutually exclusive classification of a case as a whole. In the illustrative cases discussed in Section 3, for example, the Dutch childcare benefits scandal generates primary observations in equity debris (families' financial and custodial harm) and legitimacy debris (cabinet resignation) simultaneously, while the accountability question—whether responsibility was ever clearly assigned—remains a separate, third observation to be scored independently against its own discriminating question.

## 5. Why AI debris is a governance problem, not only an ethics problem

AI debris offers three different perspectives on AI governance.

### *5.1 From model-centric to institution-centric governance*

Governance discussions often highlight model characteristics like fairness, explainability, transparency, accuracy, and robustness (Barocas et al., 2019). AI debris shifts the focus to the institution, asking*: what does the organisation become after AI adoption and withdrawal?*

### *5.2 From compliance to resilience*

Governance structures frequently prioritise documentation and procedural adherence. However, unresolved issues suggest that compliance alone may not avert long-term institutional risk. This aligns with worries that audits and governance practices can devolve into mere performances—representations of accountability devoid of meaningful structural improvements (Raji et al., 2020).

### *5.3 From deployment to lifecycle governance*

AI governance is typically viewed as a lifecycle, but the "end-of-life" stage is often overlooked. Effective governance should include withdrawal obligations, meaning not only stopping the model but also addressing its impacts.

## 6. Decommissioning Protocols: Governing AI Withdrawal and Residual Risk

The AI debris concept underscores that withdrawing AI systems entails more than technical shutdowns; it requires careful governance. Decommissioning is a crucial part of the AI lifecycle because system-level risks can persist even after a model is removed or a contract is terminated. AI withdrawal does not automatically resolve issues such as decision-making biases, workflow dependencies, or legitimacy concerns. Therefore, organisations need formal AI withdrawal protocols that focus on traceability, remediation, accountability, and institutional learning, rather than treating withdrawal as merely an endpoint.

To operationalise AI debris effectively post-deployment, Table 1 outlines the AI Debris Decommissioning Protocol (AIDP), detailing specific actions, required evidence, and governance references. Paper compliance means having governance documents and oversight forms in place but lacking the operational capacity, clear accountability, and institutional legitimacy to act effectively.

This protocol is applicable to both the public and private sectors, particularly in critical areas such as recruitment, credit allocation, fraud detection, public service triage, and early warning systems. Notably, the AIDP recognises that withdrawals often occur under chaotic conditions and prepares for triggers such as operational failures, bias incidents, regulatory actions, reputational issues, vendor terminations, or loss of maintenance capabilities.

*6.1 Feasibility and incentive-compatibility of the AIDP*

The AIDP as specified in Table 1 sits in tension with the paper's own account of accountability debris. Section 4.2 argues that organisations facing a failed or withdrawn AI system have a structural incentive to avoid assigning blame, precisely to limit legal exposure and reputational damage (Kroll, 2017)—yet Step 9 of the AIDP asks organisations to do exactly this: formally name a responsible officer and document accountability for the system's residual risk. Taken at face value, the protocol asks institutions to voluntarily produce the evidence most likely to be used against them, which is not a credible expectation absent some change to the surrounding incentive structure. Three considerations bear on when adoption becomes plausible rather than aspirational. First, voluntary self-imposed adoption is unlikely to scale on its own, since any single organisation that documents its own failure candidly bears a first-mover cost that competitors who do not adopt the AIDP avoid—a collective-action problem common to disclosure regimes generally (Sufrin & Olson, 1966). Second, adoption becomes more plausible where it is externally imposed rather than internally initiated: written into procurement contracts by large institutional buyers (banks, government agencies, health systems) as a condition of vendor engagement or mandated by regulators as part of licensing or audit requirements, in the way NIST's AI RMF already functions as a quasi-mandatory reference point for regulated sectors (NIST, 2023). Under external imposition, the incentive shifts from "why would we document our own failure" to "we lose the contract or licence if we do not," which is a materially different calculus. Third, safety-critical industries offer a relevant precedent for the mechanism that makes candid post-incident disclosure sustainable: non-punitive, "just culture" reporting systems that legally or institutionally separate reporting a failure from being penalised for it (Reason, 1997).

**Table 2.** AI Debris Decommissioning Protocol (AIDP).

This table details a protocol for AI withdrawal post-deployment, including governance purpose, auditable evidence, and reference points. It helps assess residual risk and integrate oversight into procurement, audits, and reporting.

| Step | AIDP Step | Purpose (why this step matters) | Minimum evidence / auditable artefacts | Key anchor reference(s) |
|---|---|---|---|---|
| 0 | Define the withdrawal trigger and decommissioning threshold | Prevents "silent abandonment" and ensures withdrawal is treated as a formal governance event rather than an informal technical shutdown. | Decommissioning trigger criteria; decision memo; risk threshold statement; signed withdrawal authorisation. | *NIST (2023); Kroll (2017)* |
| 1 | Freeze the decision footprint | Preserves traceability of the system's outputs and downstream impacts to prevent post-withdrawal ambiguity and blame diffusion. | Model outputs archive, decision logs, alert history, decision-to-action timeline, and data snapshots. | *Kroll (2017); Papagiannidis et al., 2025* |
| 2 | Document system scope and operational boundaries | Clarifies what the system was designed to do vs. what it was used for, reducing post-hoc rationalisation and institutional drift. | System card/model card; intended-use statement; boundary conditions; deployment map; user roles list. | *NIST (2023); Kroll (2017)* |
| 3 | Conduct an incident and near-miss review | Identifies operational failures that persist as institutional residue even after the system is withdrawn. | Incident register; near-miss log; false alarm/missed event metrics; root-cause summary. | *Papagiannidis et al., 2025; Raji et al. (2020)* |
| 4 | Audit data integrity and contamination risk | Detects whether the system altered institutional data streams, labels, or categories in ways that persist post-withdrawal. | Data lineage report; bias audit notes; dataset version history; contamination assessment; data retention plan. | *Raji et al. (2020); NIST (2023)* |
| 5 | Assess procedural debris (workflow persistence) | Identifies AI-shaped routines that continue after withdrawal (e.g., escalation shortcuts, dashboard dependency, altered SOPs). | SOP comparison (pre/post AI); workflow maps; escalation pathway analysis; override practice review. | *Kroll (2017); Papagiannidis et al., 2025* |
| 6 | Assess capability debris (judgment displacement) | Evaluates whether human judgment capacity and institutional expertise were degraded, even if workflows are formally restored. | Training gaps assessment; operator interviews; decision-quality review; deskilling indicators; staffing capability matrix. | *Papagiannidis et al., 2025* |
| 7 | Assess legitimacy and equity debris | Determines whether the system produced persistent trust erosion, exclusion, or perceived unfairness that survives system removal. | Complaint/grievance records; stakeholder feedback; inclusion audit; distributional impact notes; access gaps. | *Haraway (1988); Lyons et al. (2021)* |
| 8 | Establish contestability and redress closure | Ensures affected parties can still challenge past decisions and seek correction or compensation after withdrawal. | Redress process; appeals channel; resolution timeline; corrective action documentation; closure report. | *Lyons et al. (2021)* |
| 9 | Assign accountability for post-withdrawal risk | Prevents accountability vacuum by formally assigning responsibility for residual risk and remediation actions. | Accountability map (RACI); named responsible officer; governance ownership statement; regulator notification if required. | *Kroll (2017); NIST (2023)* |
| 10 | Implement corrective actions and remediation | Converts evaluation into institutional repair: revising SOPs, retraining staff, restoring data integrity, and repairing service gaps. | Corrective action register; implementation status; SOP updates; retraining evidence; system replacement plan. | *NIST (2023); Papagiannidis et al., 2025* |
| 11 | Conduct legitimacy repair and public communication | Restores institutional credibility by distinguishing governance repair from PR and ensuring transparency about failures and changes. | Communication record; public notice (where applicable); stakeholder briefings; updated accountability statement. | *Haraway (1988); Kroll (2017)* |
| 12 | Archive and institutionalise learning | Prevents recurrence by embedding lessons into procurement, audit, and governance processes. | After-action review (AAR); lessons-learned memo; procurement clauses update; governance checklist update. | *NIST (2023); Raji et al. (2020)* |

The AIDP should be included in procurement contracts, internal audits, and regulatory reports, treating AI withdrawal as a formal governance process instead of an informal technical shutdown. This approach not only helps address failures but also ensures responsible scaling and institutional resilience from the outset.

Aviation's incident-reporting regimes function only because reporting is protected from certain liability consequences; an AIDP without an analogous safe-harbour provision for Steps 1, 3, and 9—freezing the decision footprint, incident review, and accountability assignment—risks being adopted only by organisations with the least to hide, which is precisely the population least in need of it. The AIDP is therefore best understood not as a voluntary governance checklist but as a regulatory and procurement instrument whose feasibility depends on external enforcement and some protection for good-faith disclosure, rather than on organisations' willingness to self-impose it.

The protocol aims for three main outcomes:

1. *Preserves Traceability:* It documents decisions and system boundaries to eliminate ambiguity and avoid blame shifting after withdrawal.
2. *Identifies and Mitigates Residual Risk:* It audits for issues like data contamination and procedural dependency, ensuring there are mechanisms for contestability and redress.
3. *Institutionalises Learning:* It promotes post-incident reviews and tracks corrective actions to prevent recurrence in future deployments.

By establishing withdrawal as a governance stage, the AIDP shifts the narrative from viewing AI failure as a simple technical breakdown to recognising it as an ongoing institutional challenge with significant impacts. This framework not only highlights post-deployment risk but also strengthens AI governance beyond mere risk management, supporting long-term institutional resilience.

## 7. Future direction

### *7.1 Toward an AI Debris Index*

Future work might create an AI Debris Index that quantifies the six domains from Section 4 using evidence-based indicators outlined by the AIDP. This would enable scoring and comparing debris levels across systems and sectors, moving beyond simple qualitative classifications.

### *7.2 Comparative cases across sectors*

Debris differs by sector. For instance, banking exhibits greater accountability debris due to reliance on vendors and regulatory complexity, whereas public services exhibit greater equity debris due to citizen exclusion. Comparative research could systematically map these patterns (Eubanks, 2018; O'Neil, 2016).

### *7.3 Governance design and institutional learning*

A third approach is to pinpoint governance structures that effectively minimise debris accumulation while still fostering innovation. This resonates with the growing focus on "responsible AI" viewed as an organisational capability rather than merely a checklist compliance (Kroll, 2017; Raji et al., 2020).

## 8. Conclusion

This paper introduces the concept of AI debris to highlight a key gap in AI governance: the misconception that decommissioning AI systems fully eliminates their influence. In fact, retired AI systems can leave lasting socio-technical impacts that influence behaviour, decision-making, accountability, and stakeholder trust. Viewing AI debris as a post-deployment risk, the paper emphasises that removing a model does not eliminate its associated risks.

Four contributions are made. First, it differentiates AI debris from related concepts—namely, technical debt and safety-critical post-incident review—by temporal placement and institutional scope. This clarification highlights the truly novel aspects of a post-withdrawal governance framework, rather than treating decommissioning residue as a renaming of existing maintenance or investigation issues. Second, it outlines a typology of six debris domains, showing how residual risk can linger through procedural dependencies, data contamination, capability displacement, legitimacy erosion, and accountability breakdown, and supplies discriminant criteria for distinguishing domains that are conceptually adjacent and easily conflated in practice. Third, the AI Debris Decommissioning Protocol (AIDP) is a governance tool that aims to improve traceability, incident review, contestability, remediation, and accountability through verifiable evidence. It discusses conditions—mainly external enforcement via procurement or regulation and protections for good-faith disclosures—that make organisations more likely to adopt rather than resist such a protocol, especially when incentives favour avoiding documented accountability. Fourth, it frames decommissioning as a measurable governance stage, allowing organisations and regulators to assess ongoing effects and avoid "paper compliance," where governance appears strong on paper but is ineffective in practice.

Two cases—the discontinued Amazon hiring tool and the Dutch childcare benefits scandal—are used to illustrate different points: the former shows how procedural, capability, and equity issues often go unreported publicly, revealing how little institutions document their algorithmic processes, while the latter demonstrates issues of legitimacy, accountability, and equity through a documented record of inquiries, redress, and investigations. Together, they show both the plausibility of AI debris and the type of evidence the AIDP aims to establish routinely.

The paper emphasises that AI governance must see withdrawal as a regulated transition needing institutional repair and learning, not just a technical step. Regulators should consider using an AIDP-style withdrawal report for high-stakes AI affecting rights, safety, or resources, likely driven through procurement and licensing rather than voluntary disclosure. Future research could study AI debris pathways, validate proposed indicators for an AI Debris Index, and examine how post-deployment governance impacts long-term trust, resilience, and responsible innovation.